\documentclass[prd,preprintnumbers,floatfix,aps,nofootinbib,notitlepage,showpacs]{revtex4-1}

\usepackage{latexsym}
\usepackage{epsfig}
\usepackage{amssymb}

\newcommand{\lp}{\left(}
\newcommand{\rp}{\right)}
\newcommand{\lb}{\left[}
\newcommand{\rb}{\right]}

\newcommand{\ba}{\begin{eqnarray}}
\newcommand{\ea}{\end{eqnarray}}
\newcommand{\be}{\begin{equation}}
\newcommand{\ee}{\end{equation}}

\newcommand{\lam}{\hat{\lambda}}
\newcommand{\ld}{\lambda}

\newcommand{\al}{\alpha}
\newcommand{\bt}{\beta}

\newcommand{\la}{\lambda}

\newcommand{\Ga}{\Gamma}

\newcommand{\F}{\mathcal{C}}
\newcommand{\T}{\mathcal{T}}
\newcommand{\R}{\mathcal{R}}

\newcommand{\LA}{\mathcal{L}}
\newcommand{\vp}{\tilde{\varphi}}
\newcommand{\vpu}{\tilde{\varphi}^{(1)}}
\newcommand{\vpd}{\tilde{\varphi}^{(2)}}

\begin{document}

\title{Unifying Einstein and Palatini gravities}

\author{Luca Amendola}
\email{l.amendola@thphys.uni-heidelberg.de}
\affiliation{Institut f\"ur Theoretische Physik, University of Heidelberg, Philosophenweg 16, 69120 Heidelberg, Germany}
\author{Kari Enqvist}
\email{kari.enqvist@helsinki.fi}
\affiliation{Physics Department, University of Helsinki, and Helsinki Institute of Physics, FIN-00014 University of Helsinki.}
\author{Tomi Koivisto}
\email{t.s.koivisto@uu.nl}
\affiliation{Institute for Theoretical Physics and Spinoza Institute, Leuvenlaan 4, 3584 CE Utrecht, The Netherlands.}

\date{\today}
\pacs{04.50.Kd,04.50.-h,04.20.Fy,11.30.-j}
\preprint{HIP-2010-26/TH}
\preprint{ITP-UU-10-37}
\preprint{SPIN-10/32}

\begin{abstract}
We consider a novel class of $f(\R)$ gravity theories where the
connection is related to the conformally scaled metric $\hat
g_{\mu\nu}=C(\R)g_{\mu\nu}$ with a scaling that depends on the
scalar curvature $\R$ only. We call them C-theories and show that
the Einstein and Palatini gravities can be obtained as special
limits. In addition, C-theories include completely new physically
distinct gravity theories even when $f(\R)=\R$.
With nonlinear $f(\R)$, C-theories interpolate and extrapolate the
Einstein and Palatini cases and may avoid some of their
conceptual and observational problems. We further show that
C-theories have a scalar-tensor formulation, which in some special
cases reduces to simple Brans-Dicke-type gravity.
If matter fields couple to the connection, the conservation laws in
C-theories are modified. The stability of perturbations about flat space
is determined by a simple condition on the lagrangian.
\end{abstract}

\maketitle

\section{Introduction}

Metric determines distances and connection the parallel transport of
vectors. In principle these are independent fields: the metric has
$D(D+1)/2$ and the connection $D^3$ degrees of freedom in $D$-dimensional spacetime. In the
general theory of relativity (GR), the connection is uniquely
determined by the metric \cite{Wald:1984rg}, but no fundamental
principle dictates such prescription. Nevertheless, even in the
presence of other connections, the metric does spontaneously
generate its Levi-Civita connection, which always has a physical
relevance since matter fields follow the geodesics given by this
connection (assuming minimal coupling to geometry). Thus we may call
$g_{\mu\nu}$ the matter metric and the corresponding connection
$\Gamma$ the \emph{matter connection}. The connection $\hat{\Gamma}$,
which determines the curvature of spacetime, is the \emph{geometric connection} that enters the gravitational lagrangian. In
general, $\Gamma\ne \hat{\Gamma}$, in which case gravity is said to
be nonmetric.

One of the earliest physical ideas utilizing nonmetricity of
connections was Weyl's conformally invariant gravity, which
postulated a gauge symmetry with respect to local changes of scales
\cite{O'Raifeartaigh:1998pk}. There the incompatibility of the
metric with the connection is characterized by a single vector
$A_\alpha$  in such a way that $\hat{\nabla}_\alpha
g_{\mu\nu}=A_{\alpha}g_{\mu\nu}$, where $\hat{\nabla}$ is the
covariant derivative with respect to the independent
connection\footnote{The gauge symmetry in Weyl's theory is then
$g_{\mu\nu}\rightarrow e^\varphi g_{\mu\nu}$, $A_\alpha\rightarrow
A_\alpha-\varphi_{,\alpha}$. This demonstrates why one needs a
nonzero $A_\alpha$ (and why one could identify it with the
electromagnetic potential).}. This turns out to be the Levi-Civita
connection of the metric $\hat{g}_{\mu\nu}$ that is related to
$g_{\mu\nu}$ by the conformal transformation $\hat{g}_{\mu\nu}=\F
g_{\mu\nu}$.

In the present paper we propose that the two connections, $\Gamma$
and $\hat{\Gamma}$, have a functional relation that depends
nontrivially upon the spacetime curvature. We assume the conformal relation
\be\label{startingpoint}
\hat{g}_{\mu\nu}=\F(\R)g_{\mu\nu}~,
\ee
where $\F$ is an arbitrary function of the Ricci curvature scalar $\R=\R[g,\hat\Gamma]$ only.
We denote the curvature related to the matter connection $\Gamma$ by $R=R[g]$ and
call the class of theories obeying the relation (\ref{startingpoint}) \emph{C-theories}.

Besides
simplicity, one reason for the choice (\ref{startingpoint}) is that there are
suggestions that nonlinear actions involving $\R$ avoid the Ostrogradski and ghost instabilities
otherwise generic in higher derivative theories
\cite{Woodard:2006nt}. Moreover, it allows us to make contact with the
$f(\R)$ theories on which there exists an extensive literature (for
reviews, see
\cite{Nojiri:2006ri,Capozziello:2007ec,Sotiriou:2008rp,Lobo:2008sg,DeFelice:2010aj}),
that has inflated during the last decade or so because of the
possibility to explain the observed acceleration by an infrared
modification of gravity. There are two popular formulations of these
theories, the metric and the Palatini, the latter employing the
variational principle regarding the metric and the connection as
independent variables. {We stress that these are physically
different theories rather than manifestations of the same theory in different
guises, as the two variational principles yield inequivalent
equations of motion (except when the action is the Einstein-Hilbert and matter
is minimally coupled to geometry). Stability and Solar system
constraints impose restrictions upon the form of $f(R)$ in the
metric formalism, and cosmological implications on the
background expansion \cite{Amendola:2006kh,Amendola:2006we} and structure formation
\cite{Tsujikawa:2007tg,delaCruzDombriz:2008cp} further constrain the potentially viable
forms into somewhat complicated functions \cite{Hu:2007nk,Starobinsky:2007hu,Miranda:2009rs}.
In the Palatini formulation, Solar system constraints are passed
whilst there are certain implications on the physics of stars
\cite{Reijonen:2009hi}. The evolution of cosmological perturbations in these
models has been shown to be incompatible with the observed large scale structures \cite{Koivisto:2005yc,Koivisto:2006ie}, see though \cite{Koivisto:2007sq}
(bounds on $f(\R)$ can be also be derived from the background expansion \cite{Amarzguioui:2005zq,Fay:2007gg}). However, recent investigations
show that the models may be relevant to inflation or bouncing cosmology \cite{Barragan:2009sq,Koivisto:2010jj,Tamanini:2010uq}.

An additional physical motivation for adopting the form (\ref{startingpoint}) is that then (and only then) the
causal structure underlying the matter and the geometric connections
is the same. The \emph{Weyl rescaling}, or conformal transformation,
changes the measure of distances but not the angles between vectors,
hence leaving the lightcones invariant. In our case the Weyl potential would read
$A_\alpha=\log{\F}_{,\alpha}$ so that the form of nonmetricity we
consider here would correspond to pure gauge.

Let us note that although we
employ the conformal rescalings, C-theory explicitly breaks
conformal invariance as  of course does GR, too (there exist recent
attempts at conformally invariant gravity, e.g.
\cite{Mannheim:1988dj}). In the fibre bundle description
\cite{Nakahara:2003nw}, the separation of the tangent space into
vertical and horizontal subspaces is given by a slightly modified
rule that can depend upon the local curvature.

Perhaps surprisingly, the rather minimalistic extension  (\ref{startingpoint}) of GR is found to lead to a
variety of new physical features in the structure of gravity. In
particular, not only the right hand but also the left hand side of
the Einstein field equations is modified. Moreover, C-theories open
a unified view on both the metric and the Palatini formulations of the $f(\R)$ theories, which
are seemingly quite detached from each other. Here they appear as two specific limits
with $\F=1$ corresponding to the metric and $\F=f'(\R)$ to the Palatini formulation.
It becomes clear
that these two formulations possess quite special features among the broader class of C-theories: in the
metric case the geometric
connection coincides identically with the matter connection, while in the Palatini case the
geometric connection becomes nondynamical\footnote{Our interpretation of the different connections is
seemingly somewhat orthogonal to the one advocated in e.g. Refs. \cite{Allemandi:2004yx,Capozziello:2008it},
where $g_{\mu\nu}$ is regarded responsible
for the causal structure and $\hat{\Gamma}$ for the geodesic structure. As mentioned above, the causal structures
are identical (even in our generalized theories since we neglect a possible disformal contribution in
the relation between the metrics here) and it is $\Gamma$ that determines the physical trajectories.}. In the
latter
case the geometric connection may thus be regarded
as only an auxiliary field one uses to generate modified field equations. Such field equations then in fact
are, at the very least, problematical due to their strange derivative structure that has been extensively
discussed in the literature \cite{Sotiriou:2008rp}. The Palatini formulation in general seems not to have a consistent formulation of the Cauchy problem \cite{LanahanTremblay:2007sg},
and moreover its solutions exhibit unpleasant singularities \cite{Barausse:2007pn}. These features are absent in the
general class of theories we consider, and can be traced to the fact that in the Palatini formulation the
independent connection is in the Einstein frame. In any other case, the conformal relation between the two connections
(or equivalently, the two metrics), has a physical role in the sense that it is a dynamical quantity consistently
with the principle of minimal action. This introduces novel kind of gravitational dynamics even in the case $f(\R)=\R$.

In addition to allowing a unified view of the metric and Palatini formulations, C-theories also contain theories that interpolate between the two.
They also include theories that extrapolate them. All of these are, in principle, physically inequivalent.

Furthermore, the presence of higher spin fields such as spinors which couple explicitly to connection, new effects
arise also in the matter sector. It is natural interpret that such fields are nonminimally coupled to the geometry,
and should feel the geometric instead of the matter connection. This implies a generalization of the equivalence
principle and leads us to define a generalized stress energy tensor (or ''the hyper stress tensor''). The
metric by itself still determines the motion of particles, but the geodesics of the matter fields can be different when the
geometric connection is allowed to vary dynamically. This is one of the testable predictions of these theories. This
can be compared with the metric-affine gauge theories of gravity \cite{Hehl:1994ue}, where spinning matter generates torsion of spacetime.
There the Palatini variational principle is at play and matter is coupled to the connection. This has been
also considered in the case of nonlinear functions $f(\R)$ \cite{Sotiriou:2006qn}.

A crucial difference in our set-up with respect to the metric-affine one is
that in our case there is {\it a priori} a metric which is the potential for the independent connection. We can make
this assumption since we are restricted to Weyl spaces, i.e. the structure group of the fibre bundle is then the
conformal subgroup of the affine group of transformations that allows considerably less freedom for the connection.
Promoting the metric that generates the independent connection instead of the latter itself into the fundamental
degree of freedom, may seem a conceptually better starting point as a connection is not a tensor field. A practical
consequence of prescription (\ref{startingpoint}) is that the resulting theory avoids the inconsistency with the projective invariance that appears in the
metric affine theories. Namely, there the gravity side is invariant under the transformations
$\hat{\Gamma}^\alpha_{\mu\nu} \rightarrow \hat{\Gamma}^\alpha_{\mu\nu}+\delta^\alpha_\nu V_\mu$ by an arbitrary
vector $V_\mu$, but the matter side in general is not. The theories we consider do not share this inherent inconsistency because of the constraint on the connection.

Our prescription (\ref{startingpoint}) is described by an action involving an infinite loop, since the connection is a function
of the curvature which is generated by the connection {\it etc ad inf}. This results in a nonlocal action\footnote{Perhaps emerging from a hypothetical
strange loop quantum gravity yet to be developed. From string field theory of the closed sector, one also expects nonlocal gravitational actions with an infinite number of derivatives. Implications of such actions are currently investigated in several contexts \cite{Aref'eva:2006et,Calcagni:2010ab,Biswas:2010zk,Barnaby:2010kx}.}, and we shall find that there is indeed an equivalent scalar tensor theory bearing some
resemblance to a localized version of a nonlocal gravity theory. We show that a local version of C-theories can be formulated by applying
lagrange multiplier that imposes the constraint (\ref{startingpoint}). It is rather trivial that the Palatini variation, when subjected to the
metricity constraint, is equivalent to the metric variation \cite{Iglesias:2007nv}. This was discussed already in Ref.\cite{Sapko:1976:LMG} and equivalence has been established even
at the quantum level \cite{Fradkin:1984ai} in the Einstein-Hilbert case. In fact lagrange multipliers in gravitation have a long history\footnote{Recent applications include vector theories like the Einstein-Aether theory \cite{Jacobson:2004ts,Zuntz:2010jp} and modified gravity with reduced degrees of freedom \cite{Capozziello:2010uv,Cai:2010zm}.}, see e.g. \cite{RevModPhys.21.497}.
In the so called constrained first order formalism they have
been applied to generalized gravitational lagrangians also taking into account nonmetricity, see \cite{Querella:1998ke} for a review.
However, in addition to the most peculiar feature of C-theories which is that 1) the nonmetricity is prescribed to be given
by curvature, the differences to all previous literature here are that 2a) in our case the constraints is imposed upon the relation of metrics
(and not of the connections) and that 2b) we do not apply the Palatini variational principle. Nevertheless, a similar conclusion to ours that the
unconstrained variation results in inconsistencies which are cured by imposing the constrains has been reached also in e.g. \cite{Cotsakis:1997c}.
Indeed, the difference 2) here may be at least classically inessential to the general idea,
though technically allows us to explore a slightly different framework where we need vary only tensors (and avoid an additional index).
We stress that the novelty of C-theories and the new insight they provide upon Einstein and Palatini gravities emerge from the simple \emph{proposal that the relation between the matter and the geometric connections is curvature-dependent}, which results formally in an infinite derivative theory\footnote{At this point it is clear to the reader that the ''C'' of C-theories may refer as well to the modified Connection, to the Conformal relation or to the Constrained variation.}.

We will formulate the action principle using a lagrangian multiplier in section \ref{sec_action}.
In section \ref{bst} it is shown that the theory can also be presented as a scalar-tensor theory, which does not
reduce to the usual ones except for some specific forms of the action or in vacuo. The field equations are then
written down in section \ref{field_eq} to analyze structure of the theory more detail. In particular, the
conservation laws are derived when matter is allowed to couple to the connection. The perturbative stability of these theories
is investigated in section \ref{stability}. Finally, before concluding in section \ref{conclusions}, we consider
how the set-up could arise from two dual actions involving nonlinear dependence upon the curvature in section
\ref{duality}.
Few details of scalar-tensor formulations of the theories
are confined to the appendices.

\section{Action}
\label{sec_action}

To specify our notation and conventions, we begin by reviewing the definitions of some curvature quantities and conformal relations between them.
The Ricci curvature scalar can be defined by as a function of both the metric and an independent connection $\hat{\Ga}$ through
\be
\R  \equiv  g^{\mu\nu}\hat{R}_{\mu\nu} \equiv g^{\mu\nu}\lp \hat{\Gamma}^\alpha_{\mu\nu , \alpha}
       - \hat{\Gamma}^\alpha_{\mu\alpha , \nu} +
\hat{\Gamma}^\alpha_{\alpha\lambda}\hat{\Gamma}^\lambda_{\mu\nu} -
\hat{\Gamma}^\alpha_{\mu\lambda}\hat{\Gamma}^\lambda_{\alpha\nu}\rp\,.\label{r_def}
\ee
The definition of the metric Ricci scalar $R$ is, as usually, the above formula unhatted.  If the independent connection is compatible with a metric, there is a $\hat{g}_{\mu\nu}$ such that
\be
\hat{\Ga}^\gamma_{\al\bt} = \frac{1}{2}\hat{g}^{\gamma\mu}\lp \hat{g}_{\al\mu,\bt}+\hat{g}_{\bt\mu,\al}-\hat{g}_{\al\bt,\mu}\rp\,.
\ee
Again this holds with hats off as well.
One can also consider the curvature scalar constructed solely from the hatted metric, which is defined by $\hat{R}   \equiv   \hat{g}^{\mu\nu}\hat{R}_{\mu\nu}$.
Furthermore, if the two metrics are related by a Weyl rescaling $\hat{g}_{\mu\nu}=\F g_{\mu\nu}$, the relation between the two connections is then
\be
\hat{\Ga}^\gamma_{\al\bt} = {\Ga}^\gamma_{\al\bt} + \lp 2\delta^\gamma_{(\alpha}\delta^\lambda_{\bt)}-g_{\al\bt}g^{\gamma\lambda} \rp\nabla_\lambda \log{\F}\,.
\ee
This fixes the form of Weyl's nonmetricity vector in a unique way. It is then straightforward to find the transformation of the Ricci tensor,
\be \label{conformal}
\hat{R}_{\mu\nu} = R_{\mu\nu} + \frac{1}{4\F^2}\lb 3\lp D-2 \rp \F_{,\mu}\F_{,\nu}-\F\nabla_\mu\F_{,\nu}-\lp D-4 \rp (\partial\F)^2 g_{\mu\nu}\rb - \frac{1}{2\F}\lb\lp D-2\rp \nabla_\mu\nabla_\nu\F + g_{\mu\nu}\Box\F\rb\,,
\ee
from which follows that the three scalar curvatures (there would be a fourth possible but we have no use for it) are related as
\be \label{r_conformal}
\R =  \F\hat{R} = R  - \frac{D-1}{4\F^2}\lb 4\F\Box\F+(D-6)(\partial\F)^2\rb\,.
\ee
 It is well known that the above relations naturally appear in $f(\R)$ gravities \cite{Magnano:1993bd}. In particular, Palatini variation of an $f(\R)$ action yields the result that the independent connection $\hat{\Ga}$ is compatible with the conformal metric $\hat{g}_{\mu\nu} = f'(\R)g_{\mu\nu}$.  The conformal factor $\F$ that relates the Jordan and the Einstein frame in the Palatini as well as metric $f(R)$ theories has the same functional dependence upon $R$, $\F=f'(R)$ (when $D=4$).

Here we consider the more general case where the connection is conformally related to the
metric by an arbitrary function $\F(\R)$. Since this function in turn depends on the connection (or
equivalently, the conformal metric that is the potential for the connection), we are lead to the feedback loop
\be \label{loop}
\R = \R\lp g,\hat{g}\lp\R(g,\hat{g}\lp\R(\dots) \rp)\rp \rp\,.
\ee
A way to realize this set-up is to introduce a langrange multiplier $\ld^{\mu\nu}$ that enforces the desired
condition on the metric $\hat{g}_{\mu\nu}$. Defining
\be \label{lambda_def}
\ld \equiv g_{\mu\nu}\ld^{\mu\nu}\,, \quad
\hat{\ld} \equiv \hat{g}_{\mu\nu}\ld^{\mu\nu}\,,
\ee
we may write the action as
\be \label{action}
S= \int d^D x \sqrt{-g} \lb f(\R) + \hat{\ld} - \F(\R)\ld + 16\pi G \mathcal{L}_m(\Psi,\hat{\nabla}_\alpha\Psi,g_{\mu\nu}) \rb\,.
\ee
Formally solving it and plugging the solution back yields an $f(\R)$ theory where $\R$ is given by (\ref{loop}).
The matter fields residing in the lagrangian function $\mathcal{L}_m$ are denoted collectively by $\Psi$. We assume
the minimal prescription generalizing lagrangians from flat to curved spacetime
as $\eta_{\mu\nu}\rightarrow g_{\mu\nu}$, $\partial_\alpha\rightarrow\hat{\nabla}_\alpha$.

\subsection{Limiting $f(\R)$ gravities}

We construct a class of theories which interpolates between the metric and Palatini gravities, let us consider a one-parameter family of functions $\F_\al(\R)$ such that $\F_0(\R)=1$ and $\F_1(\R)=(f'(\R))^\frac{2}{D-2}$.
A simple example is $\F_\al(\R)=(f'(\R))^\frac{2\al}{D-2}$. One may also employ the linear form
\be \label{alpha}
\F_\al(\R) = \al (f'(\R))^\frac{2}{D-2} + 1 - \al\,.
\ee
With such a choice, the conformal relation between the metrics interpolates between the metric and the Palatini cases when $\al \in [0,1]$.
Now it is crucial to observe that the lagrangian multiplier terms also contribute to the dynamics, in addition to imposing the
conformality of the metrics. In the case $\alpha=1$ the presence of $\lambda_{\mu\nu}$ modifies the field
equations in precisely the correct way to guarantee the dynamical equivalence with the metric $f(R)$ theory. In the limit $\alpha=1$ however,
we obtain additional dynamics. Thus this C-theory augments the Palatini models in such a way that it contains all the solutions of the usual Palatini theories, but also an additional degree of freedom. This cures the problems like the ill-defined Cauchy formulation and the appearance of curvature singularities in simple solutions, since they all stem from the lack of dynamics. Of course all solutions of the C-theory versions of Palatini-$f(\R)$ gravity retain the same functional dependence relating the two metrics $\hat{g}_{\mu\nu} = (f'(\R))^{\frac{2}{D-2}}g_{\mu\nu}$.

It is however illustrative to use a slightly different parameterization of the action which yields precisely the usual Palatini-$f(\R)$ in the limit $\alpha=1$. For this purpose we multiply the constraint term by a constant which is otherwise an irrelevant rescaling of the lagrangian multiplier, but ensures that the multiplier doesn't introduce additional dynamics in the case $\alpha=1$:
\be \label{action_a}
S_\alpha= \int d^D x \sqrt{-g} \lb f(\R) + \lp 1-\alpha\rp \lp \hat{\ld} - \F_\al(\R)\ld\rp + 16\pi G \mathcal{L}_m(\Psi,\hat{\nabla}_\alpha\Psi,g_{\mu\nu}) \rb\,.
\ee
To recapitulate, this action interpolates between the metric and the Palatini theories when $\al \in [0,1]$. Without the term $(1-\alpha)$, the only difference is that at $\al=1$ we obtain the C-theory version of Palatini gravity sharing the solutions of the usual one but not its pathologies.

It may be useful to observe that the action (\ref{action}) can have a mapping to $\hat{f}(\hat{R})$ gravity in the $\F$-frame (the hatted frame). Observe that $\R/\F(\R)=\hat{R}$ and suppose there is a solution $\R=r(\hat{R})$. Then one may write the action in the $\F$-frame as
\be \label{hatfr}
S = \int d^D x\sqrt{-\hat{g}}\hat{f}(\hat{R}) + 16\pi G S_m(\Psi,\hat{\nabla}_\alpha\Psi,\hat{g}_{\mu\nu}/\F(r(\hat{R}))\,,
\ee
where
\be
\hat{f}(\hat{R}) = \F^{-\frac{D}{2}}\lp r(\hat{R})\rp f\lp\hat{R}\rp\,.
\ee
As an example, the power-law case $f(\R) \sim \R^n$, $\F \sim \R^m$ is $\hat{f}(\hat{R}) \sim \hat{R}^\frac{n-\frac{D}{2}m}{1-m}$ gravity in the $\F$-frame. We look at the vacuum of this example in the Jordan and Einstein frames in subsection \ref{einstein}.
Here we should emphasize that the action (\ref{action}) is more general than $\hat{f}(\hat{R})$ gravity (\ref{hatfr}) at least because there doesn't always exist a solution to the equation $\R/\F(\R)-\hat{R}=0$, the simplest degenerate case being a linear $\F(\R)$. In addition, this equation in general has multiple roots, resulting in an interesting picture where the same spacetime in different regions is effectively described by a different $\hat{f}(\hat{R})$ theory. In Palatini-$f(\R)$ models an analogous situation can occur at a less fundamental level when the algebraic trace equation has multiple roots so that the solutions of the same theory can interpolate between regions of space with a different cosmological constant.

\section{(Bi)scalar-tensor formulations}
\label{bst}
As is well known, by a suitable transformation $f(\R)$ theories can be expressed as scalar-tensor theories
with
\be
\mathcal{L}_g = \phi R - \frac{\omega_{BD}}{\phi}\lp\partial\phi\rp^2 + f({f'}^{-1}(\phi))-\phi {f'}^{-1}(\phi)\,,
\ee
where the Brans-Dicke coupling parameter $\omega_{BD}=0$ and $\omega_{BD}=-(D-2)/(D-1)$ respectively for the metric and Palatini cases\footnote{
Brans-Dicke theories with the shifted $\omega_{BD}$ parameter $\omega_{BD} \rightarrow (D-1)\phi/(D-2)(\phi-\Omega_A)$ generalize the two versions of $f(\R)$ theories. For this one-parameter class of theories, the field is an algebraic function $\phi=\phi((D-2)\Omega_A R + 16\pi T)$. The case $\Omega_A=0$ is equivalent to $f(\R)$ theory, $\Omega_A=1$ corresponds to $R+f(\R)$ theory, and in the limit $\Omega_A\rightarrow \infty$ one recovers $f(R)$ gravity \cite{Koivisto:2009jn}. The algebraic property singles out uniquely the interpolation within the restricted (almost)-Brans-Dicke context. However, we shall see that the C-theories are not in general confined to the Brans-Dicke class of scalar-tensor theories.}. Such an approach works also for the C-theories.
Adding yet a scalar lagrangian multiplier $\xi$ and an auxiliary field $\phi$, we may write a gravitational
lagrangian density equivalent to the one appearing in (\ref{action}) as
\be \label{rename}
\mathcal{L}_g = f(\phi) + \hat{\ld} - \F(\phi)\ld + \xi(\R-\phi)\,.
\ee
Its algebraic equation of motion allows to eliminate the field $\phi$ in terms of the two fields $\xi$
and $\ld$ as:
\be \label{phi}
f'(\phi)-\ld \F'(\phi)-\xi=0 \quad \Rightarrow \quad \phi=\phi(\xi,\ld)\,.
\ee
It follows that
\be
\mathcal{L}_g = f(\phi(\xi,\ld))+\hat{\ld} - \F(\phi(\xi,\ld))\ld + \xi\R - \phi(\xi,\ld)\xi\,.
\ee
We may then vary with respect to $\ld_{\mu\nu}$ to obtain (suppressing the arguments of functions)
\be
\hat{g}_{\mu\nu} = \lp \F - \frac{\partial f}{\partial \ld} + \ld \frac{\partial \F}{\partial \ld} +
\xi\frac{\partial \phi}{\partial \ld} \rp g_{\mu\nu}\,.
\ee
By virtue of Eq.(\ref{phi}), the last three terms vanish in the right hand side. We may then eliminate the
hatted metric from the lagrangian. Plugging (\ref{r_conformal}) into the action yields, after a partial integration,
\be \label{bist}
\mathcal{L}_g = \xi R + \frac{D-1}{4\F^2}\lb 4\F g^{\mu\nu} \xi_{,\mu}\F_{,\nu} - \lp D-2\rp \xi\lp\partial\F\rp^2 \rb
+ f(\phi(\xi,\ld)) -\xi\phi(\xi,\ld)\,,
\ee
where $\F$ is understood as a function of the two scalar fields $\F=\F(\phi(\xi,\ld))$. In case one is concerned about neglecting
the tensor structure in $\ld=g_{\mu\nu}\ld^{\mu\nu}$, note that the equation of motion for $\ld_{\mu\nu}$
is
\be
0=\frac{\delta\lp\sqrt{-g}\LA_g\rp}{\delta \ld^{\mu\nu}}
=\sqrt{-g}\lb \frac{\partial \LA_g}{\partial \la^{\mu\nu}}-\nabla_\al \frac{\partial
\LA_g}{\partial\lp\nabla_\al\la^{\mu\nu}\rp}\rb
=\frac{\delta\lp\sqrt{-g}\LA_g\rp}{\delta \ld}g_{\mu\nu}\,,
\ee
and therefore equivalent with the equation of motion for $\ld$. Since we can write
$\nabla_\al\ld=g_{\mu\nu}\nabla_\al \ld^{\mu\nu}$, it follows by an analogous argument that the tensor structure
in $\ld$ does not contribute to the field equations of the metric.

We should then check that the parameterization (\ref{alpha}) for a given function $f$ yields
the Brans-Dicke theories that correspond to the metric and Palatini-$f(\R)$ in the relevant limits.
In the case $\alpha=0$ the kinetic term in (\ref{bist}) is trivial, and we obtain a Brans-Dicke theory with the coupling parameter $\omega_{BD}=0$.
This corresponds to a metric $f(R)$ theory. When
$\alpha=1$, we have the solution $\ld=0$, and then we can infer from (\ref{phi}) that $\F=\xi^{2/(D-2)}$. In this
case (\ref{bist}) reduces to a Brans-Dicke theory with the coupling parameter $\omega_{BD}=-(D-1)/(D-2)$, which
corresponds to the Palatini version of an $f(\R)$ theory.

In passing we mention that a nonlocal $f(R/\Box)$ action \cite{Deser:2007jk} can also be written as a biscalar-tensor theory\footnote{However, there are subtleties in this localization to be taken into account when mapping the solutions in the resulting theory to the original starting point as remarked also in Refs.\cite{Koshelev:2008ie,Deffayet:2009ca}.} \cite{Nojiri:2007uq,Koivisto:2008xfa}.
In that case the scalar fields are massless and the kinetic term, though involves mixing, is of a different form than in our (\ref{bist}) (this may be changed by considering more general operators \cite{Koivisto:2008dh}, for example $\Delta=\nabla_\mu h(\phi) \nabla^\mu$ and adding additional term into the action $Rf(R/\Delta)\rightarrow Rf(R/\Delta) + V(R/\Delta)$).

\subsection{Einstein frame}
\label{einstein}

We consider the lagrangian (\ref{bist}) in terms of the pair of fields $(\xi,\phi)$ instead of $(\xi,\lambda)$ as it simplifies some formulas. In appendix \ref{app_xl} we consider the equivalent
system in terms of the other pair of fields. The Einstein conformal frame, denoted by a star, is reached by the conformal transformation $g^*_{\mu\nu}=(\xi/\xi_0)^\frac{2}{D-2}g_{\mu\nu}$. We obtain
\be
S^* = \int d^D x \sqrt{-g^*}\lb R^* -2\gamma_{ab}(\varphi^c){g^*}^{\mu\nu}\varphi^a_{,\mu}\varphi^b_{,\nu}
- 4B(\varphi^c)
+ 16\pi G (\xi/\xi_0)^\frac{-D}{D-2}\mathcal{L}_m(\Psi,(\xi/\xi_0)^{\frac{-2}{D-2}}g^*_{\mu\nu})\rb\,.
\ee
We are employing the notation of Ref.\cite{Damour:1992we} and the fields are denoted as $\varphi^{(1)}=\xi$, $\varphi^{(2)}=\phi$. The field $\xi$ was rescaled $\xi \rightarrow \xi/\xi_0$ so that all the fields have the dimension mass squared. The potential is
\be
B(\xi,\phi) = \frac{(\xi_0/\xi)^{\frac{D}{D-2}}}{4}\lb   \frac{\xi}{\xi_0}\phi - f(\phi)\rb\,,
\ee
and the components of the field space metric $\gamma_{ab}(\xi,\phi)$ defining the nonlinear sigma interaction is
\ba
\gamma_{\xi\xi}(\xi,\phi) & = & \frac{D-1}{2(D-2)\xi^2}\,, \nonumber \\
\gamma_{\xi\phi}(\xi,\phi) & = & \gamma_{\phi\xi}(\xi,\phi)=\frac{-(D-1)\F'(\phi)}{4\F(\phi)\xi}\,, \nonumber \\
\gamma_{\phi\phi}(\xi,\phi) & = & \frac{(D-1)(D-2){\F'}^2(\phi)}{8\F^2(\phi)}\,. \label{gamma_p}
\ea
Here again $\phi=\phi(\xi,\lambda)$ as given by (\ref{phi}). However, it turns out that the metric $\gamma_{ab}(\xi,\phi)$ is degenerate. One of the eigenvalues of the matrix defined by
(\ref{gamma_p}) vanishes identically,
\ba
\lambda^{(1)}_\gamma & = & 0\,, \\ \nonumber
\lambda^{(2)}_\gamma & = & \frac{D-1}{8(D-2)} \lp \frac{(D-2)^2{\F'}^2(\phi)}{\F^2(\phi)}+\frac{4}{\xi^2}\rp\,.
\ea
Therefore we cannot invert $\gamma_{ab}(\xi,\phi)$ and straightforwardly implement the results of \cite{Damour:1992we} to analyze the PPN limit. This implies that there is only one additional propagating scalar degree of freedom compared to GR. That agrees with the following section where we find that the scalar $\lambda$ can be eliminated in terms of $\R$ and $T$, while $\R$ needs, in general, to be solved from a dynamical equation.
The eigenmodes corresponding to the eigenvalues $\lambda^{(1)}_\gamma$ and $\lambda^{(2)}_\gamma$ can be solved from, respectively
\ba
d\tilde{\varphi}^{(1)} & = & \frac{(D-2)\F'(\phi)\xi}{2\F(\phi)}d\xi + d\phi\,, \label{varphi1} \\
d\tilde{\varphi}^{(2)} & = & \frac{-2\F(\phi)}{(D-2)\F'(\phi)\xi}d\xi + d\phi\,. \label{varphi2}
\ea
One easily sees that only in the case of an exponential $\F$ (\ref{varphi1},\ref{varphi2}) are two exact
differentials. We consider this specific case in the appendix \ref{exponential}.

In general we can decouple the kinetic term from the other field even if it is not orthogonal to the
dynamical field in the field space. To this effect, we introduce
\be
\psi = (D-2)\log{\F(\phi)} - 2\log{\frac{\xi}{\xi_0}}\,.
\ee
The action is then rewritten as
\be
S^* = \int d^D x \sqrt{-g^*}\lb R^* -\frac{D-1}{4(D-2)}{g^*}^{\mu\nu}\psi_{,\mu}\psi_{,\nu}-
4B(\psi,\xi)\rb
+ 16\pi G S_m(\Psi,(\xi/\xi_0)^{\frac{-2}{D-2}}g^*_{\mu\nu})\,,
\ee
where
\be \label{psi_action}
B(\psi,\xi)= \frac{(\xi_0/\xi)^{\frac{D}{D-2}}}{4}\lb \frac{\xi}{\xi_0}\phi - f(\phi)\rb\,, \quad \phi \equiv
\F^{-1}\lb
\lp\frac{e^\psi\xi^2}{\xi_0^2}\rp^\frac{1}{D-2}\rb\,.
\ee
By varying with respect to $\xi$, we obtain
\be \label{xi_eom}
\lp\frac{\xi_0}{\xi}\rp^\frac{D}{D-2}\lp D f(\phi) - 2\phi\frac{\xi}{\xi_0}\rp +
\frac{2e^\frac{\psi}{D-2}}{\F'(\phi)}\lp 1 - \frac{\xi_0f'(\phi)}{\xi}\rp = 16\pi G T\,.
\ee
The field $\phi=\phi(\psi,\xi)$ again as in Eq.(\ref{psi_action}), and $T$ is the trace of the energy-momentum tensor defined as
\be \label{set}
T_{\mu\nu} = \frac{-2}{\sqrt{-g}}\frac{\delta \lp \sqrt{-g}\mathcal{L}_m \rp}{\delta g^{\mu\nu}}\,.
\ee
We are assuming matter is coupled minimally in the Jordan frame.
Solving the auxiliary field from a constraint facilitates the analysis of the field equations. However,
eliminating the auxiliary field at the level of action in general results in an unwieldy formulation of the
theory, because of the coupling to the matter trace in (\ref{xi_eom}). This in general would couple nonminimally to both
the dynamical scalar and to gravity in the Jordan frame.

\subsection{Power-law model in vacuum}

In vacuum or in the presence of only conformal matter fields the lagrangian can be written more conveniently in terms of a single field. When $T=0$, the field $\xi$ is given by
\be \label{xi}
\frac{\xi}{\xi_0} = \frac{\F(\phi)f'(\phi)-\frac{D}{2}\F'(\phi)f(\phi)}{\F(\phi)-\F'(\phi)\phi}\,,
\ee
where $\phi$ is regarded as the shorthand for a function of $(\psi,\xi)$ as in
Eq.(\ref{psi_action}) above.
To illustrate the system with an exactly solvable an example, we consider the power-law class of models specified by the two exponents $n$ and $m$,
\be
f(\R) = f_0\R^n\,, \quad \F(\R) = c_0\R^m\,.
\ee
Here $f_0$ and $c_0$ are constants with the appropriate dimension. We obtain that
\ba \label{phi1}
\phi & = & \frac{k_0^\frac{2}{m(D-2)}}{c^\frac{1}{m}}\exp\lp\frac{\psi}{m(D-2)-2(n-1)}\rp\,, \nonumber \\
\frac{\xi}{\xi_0} & = & k_0 \exp\lp \frac{(n-1)\psi}{m(D-2)-2(n-1)}\rp\,, \quad k_0 \equiv
\lp \frac{f_0(n-\frac{D}{2}m)}{c_0^\frac{n-1}{m}(1-m)} \rp^\frac{m(D-2)}{m(D-2)-2(n-1)}\,.
\ea
One may substitute this back into the action, transform back into the Jordan frame and by a field redefinition of $\psi$ reintroduce the dimensionless $\xi$ to put the Brans-Dicke theory into its canonical form. An equivalent result is obtained by just substituting $\F=\F(\xi)$ into the action (\ref{bist}) which then becomes
\be \label{bd_action}
S = \int d^Dx\sqrt{-g}\lb \xi R - \frac{\omega_{BD}}{\xi}\lp\partial\xi\rp^2 - V_0 \xi^\frac{n}{n-1}\rb\,,
\ee
where
\ba
\omega_{BD}  & = & -\frac{m(D-1)}{4(n-1)^2}\lb 4(n-1)-m(D-2)\rb\,, \\
V_0 & = & \lp n+m-\frac{D}{2}m-1\rp \lb \frac{1-m}{(n-\frac{D}{2}m)^n f_0}\rb^\frac{1}{n-1} \label{potential}\,.
\ea
When $m=0$ the Brans-Dicke parameter $\omega_{BD}$ vanishes, as it should since $m=0$ is a metric $f(R)$ model.
In the case $m=4(n-1)/(D-2)$ there is also a mapping to a metric $f(R)$ model. The Palatini case $m=(n-1)/(\frac{D}{2}-1)$, which yields $\omega_{BD}=-(D-1)/(D-2)$ as expected. The potential (\ref{potential}) vanishes then, reflecting the fact that the trace equation in Palatini gravity has a zero cosmological constant solution in vacuum. An alternative way to look at this is to note that in the $\F$-frame this is equivalent to Einstein-Hilbert gravity, recall the consideration following (\ref{hatfr}). We note also, from Eq. (\ref{phi1}), that the field $\xi$ vanishes when $n=\frac{D}{2}m$. The reason behind this is that then the model is solely the cosmological constant in the $\F$-frame (without the Einstein-Hilbert term). Finally, from the relation (\ref{xi}) we make the interesting observation that regardless of the form of $f(\R)$, theories with linear relation $\F(\R) \sim \R$ have the special property that the field $\R$ is an algebraic function (of matter fields in general). The theory can then nevertheless be consistent and dynamical. However, the particular lagrangian $m=1$, $n=D/2$ is degenerate, because this coincides with the Palatini limit. The degeneracy of the quadratic Palatini theory has been known since long \cite{0305-4470-12-8-017}.

 To close this section, let us clarify that in general the C-theories do not reduce to simple Brans-Dicke gravity even in vacuum. The general form of the kinetic term there is such that the constant $\omega_{BD}$ in (\ref{bd_action}) is replaced by the function
\be
\omega(\xi) = \frac{(D-1) \F'{^2} (2 \F f'-D  \F'f) \lb D^2 {\F'}^3f-2 D {\F'}^2( \F'f+3 \F f') +4 \F
   \left(3\F'^2f'+2 \F \F' f'' -2 \F \F'' (f'-\xi )\right)\rb}{4 \F^2 \lb (D-2)
   {\F'}^2f'-2 \F  {\F}'f''+2 \F \F'' (f'-\xi )\rb^2 }\,,
\ee
where $f=f(\phi(\xi))$  and $\F=\F(\phi(\xi))$ such that $\phi(\xi)$ solves Eq. (\ref{xi}). Hence, in general the coupling in C-theories is nonlinear.

\begin{center}
\begin{table}
\begin{tabular}{|c||c|c|c| }
\hline
 C-theory   & $f(\R) \sim \R$  & $f(\R) \sim \R^{\frac{D}{2}}$ & general $f(\R)$\\
\hline \hline $\F(\R) \sim 1$ & GR & quadratic gravity & metric $f(R)$ \\
\hline $\F(\R)\sim \R$   & ? & degenerate & $\R=\R(T)$ \\
\hline $\F(\R) \sim {f'}^{\frac{2}{D-2}}(\R)$ & GR & degenerate & Palatini-$f(\R)$ \\
\hline
\end{tabular}
\caption{Some exceptional cases of C-theories which reduce to previously known or degenerate theories.}
\end{table}
\end{center}

\section{Field equations}
\label{field_eq}

Let us first assume that matter is minimally coupled to the metric $g_{\mu\nu}$.
By varying (\ref{action}) with respect to the three tensor fields $g^{\mu\nu}$, $\hat{g}_{\mu\nu}$ and
$\ld^{\mu\nu}$ we obtain, respectively, the following equations of motion.
\ba
8\pi G T_{\mu\nu} & = &
\lb f'(\R) - \ld\F'(\R)\rb\hat{R}_{\mu\nu} - \frac{1}{2}\lb f(\R) + \lam - \F(\R)\ld \rb g_{\mu\nu}
+ \F(\R)\ld_{\mu\nu} \,, \label{geom} \\
\la^{\mu\nu} & = &
\frac{1}{2}\sqrt{\frac{\hat{g}}{g}}\lp
\hat{g}^{\mu\nu}\delta^\gamma_\alpha\delta^\rho_\beta
+\hat{g}^{\gamma\rho}\delta^{\mu}_\alpha\delta^\nu_\beta
-2\hat{g}^{\rho(\mu}\delta^{\nu)}_\bt\delta^{\gamma}_\al
\rp
\hat{\nabla}_{\gamma}\hat{\nabla}_\rho \lb\sqrt{\frac{g}{\hat{g}}}\lp f'(\R) - \lambda \F'(\R) \rp g^{\al\bt}\rb \,, \label{hateom} \\
\hat{g}_{\mu\nu} & = & \F(\R)g_{\mu\nu}\,. \label{leom}
\ea
We denote $\lambda_{\mu\nu}=g_{\mu\alpha}g_{\nu\beta}\lambda^{\alpha\beta}$, and the stress energy was defined in (\ref{set}).
By imposing the constraint (\ref{leom}), the pair of equations (\ref{geom},\ref{hateom}) becomes
\ba
8\pi G T_{\mu\nu} & = &
\lb f'(\R) - \ld\F'(\R)\rb\hat{R}_{\mu\nu} - \frac{1}{2}f(\R)g_{\mu\nu}
+ \F(\R)\ld_{\mu\nu} \,, \label{geom2} \\
\la_{\mu\nu} & = & \F(\R)^{\frac{D-4}{2}}\lp \hat{g}_{\mu\nu}\hat{\Box}-\hat{\nabla}_\mu\hat{\nabla}_\nu\rp
\lb \F(\R)^{\frac{2-D}{2}}\lp f'(\R)-\lambda\F'(\R)\rp\rb\,. \label{hateom2}
\ea
The second derivatives of $\R$ are contained in $\hat{R}_{\mu\nu}$ and $\lambda_{\mu\nu}$. Thus the general theories are nonlinear in second derivatives of the curvature scalar, unlike the conventional fourth order gravity.

At this point, it is easy to see that we obtain the correct field equations in the appropriate limits of the
parameterization (\ref{alpha}). When $\al=0$, $\F(\R)=1$, $\R=R$ and thus we obtain the field equations for a
metric $f(R)$ theory:
\be
\al=0\, : \quad
f'(R)R_{\mu\nu} - \frac{1}{2}f(R)g_{\mu\nu} + \lp g_{\mu\nu}\Box-\nabla_\mu\nabla_\nu\rp f'(R) = 8\pi G
T_{\mu\nu}\,. \label{m_eom}
\ee
In the C-theory version of $\al=1$ parameterization, one sees that $\ld_{\mu\nu}=0$ is a solution to (\ref{hateom2}). If we consider the action (\ref{action_a}), $\ld_{\mu\nu}$ does not appear in the field equations in the first place, but the conformal relation follows as a solution to the equation (\ref{hateom}). In either case, we obtain
\be
\al=1\, : \quad f'(\R)\hat{R}_{\mu\nu} - \frac{1}{2}f(\R)g_{\mu\nu} = 8\pi G T_{\mu\nu}\,. \label{p_eom}
\ee
This is the field equation for Palatini-$f(\R)$ gravity. By using the trace of this equation,
$f'(\R)\R-Df/2=8\pi G T$ we can express the scalar curvature as an algebraic function of the matter stress energy
trace, $\R=\R(T)$.

In the general case, the structure of the theory is as follows. There are two scalar degrees of freedom $\lambda$ and $\R$ corresponding to
to the scalar fields $\lambda$ and $\xi$ in our formulation (\ref{bist}).
The field equation
(\ref{geom2}) can be written purely in terms of the metric and the scalar curvature by using the conformal
relation (\ref{conformal}) for the Ricci tensor and rewriting (\ref{hateom}) as
\be
\ld_{\mu\nu} =
\F(\R)^{\frac{D-4}{2}}\lp {g}_{\mu\nu}{\Box} - {\nabla}_\mu{\nabla}_\nu\rp S(\R,T)
-\F(\R)^{\frac{D-6}{2}}\lb (D+1)g_{\mu\nu}g^{\al\bt}\F(\R)_{,\al}S(\R,T)_{,\bt}-2\F(\R)_{(,\mu}S(\R,T)_{,\nu)}\rb\,.
\label{hateom3}
\ee
We used the short-hand notation $S(\R,T)=\F(\R)^{\frac{2-D}{2}}\lp f'(\R) -\lambda \F'(\R)\rp$. The reason is that using the trace of the field equation
(\ref{geom2})
\be
\ld\lp \F'(\R)\R -\F(\R) \rp - f'(\R)\R + \frac{D}{2}f(\R) + 8\pi G T = 0\,, \label{r_eom}
\ee
we may write the function $S(\R,T)$ as
\be \label{S}
S(\R,T)=C^\frac{2-D}{2}(\R)\lb f'(\R)+\frac{f'(\R)\R-\frac{D}{2}f(\R)-8\pi G T}{\F(\R)-\F'(\R)\R}\F'(\R)\rb\,.
\ee
It is probably more convenient to use this constraint in practise than to solve the dynamical equation for the field $\lambda$ one obtains by taking the trace of (\ref{hateom3}):
\be \label{lambda}
\ld =
(D-1)\F(\R)^{\frac{D-6}{2}}\lb \F(\R){\Box}S(\R,\lambda) -
(D+2)g^{\al\bt}\F(\R)_{,\al}S(\R,\lambda)_{,\bt}\rb\,.
\ee
This can be compared with the biscalar-tensor formulation in section \ref{einstein} where we found that the
auxiliary scalar can be eliminated in terms of the other scalar field and the matter trace.

The trace (\ref{r_eom}) appears also as an equation of motion for the scalar curvature.
Notice however, that the field equations contain second derivatives of the metric and second derivatives of the
curvature scalar $\R$. The trace doesn't introduce additional information and thus (\ref{r_eom}) and
(\ref{geom2}) cannot determine the evolution of the system in full generality. An independent piece of
knowledge is provided by the conformal relation between $R$ and $\R$ which results in the evolution equation for $\R$,
\be \label{r_evol}
\Box \F(\R) + \frac{(D-6)}{4\F(\R)}\lp\partial \F(\R)\rp^2 = \frac{\F(\R)}{D-1}\lp R -\R \rp\,.
\ee
Hence, to generate solutions one has to consider the second order dynamical equation (\ref{r_evol}) for
the scalar curvature $\R$, coupled with
the field equations for the metric (\ref{geom2}). In the Palatini limit $\alpha\rightarrow 1$ of the action (\ref{action_a}), (\ref{r_eom})
reduces to a constraint and the scalar curvature determining connection cannot settle itself dynamically to minimize the
action. It is easy to see why the formulation of the Cauchy problem is spoiled in this limit, since it is not continuous but a degree of freedom disappears with the lagrangian multiplier. In the C-theory version this doesn't occur.

We note that while $\al=0$ and $\al=1$ are very special points in the theory space, there seems to be nothing
particular in the limit $f(\R)=\R$. This applies in the case of any nontrivial $\F(\R)$, but for metric $f(R)$ actions it is well known that the linear case is a special limit where the theory reduces to second order. However, linear $\F\sim\R$ has the special property that the curvature scalar has the same functional relation $\R=\R(T)$ as in the Palatini theories, as one immediately sees from (\ref{r_eom}). Then $f(\R)=R^\frac{D}{2}$ is a special case which becomes doubly degenerate in the sense that there also the trace of the $f(\R)$-part disappears. It is interesting to note that barring these special cases, in general one has a dynamical gravity theory even in the case $f(\R)=-2\Lambda$, where in fact $\Lambda$ can vanish\footnote{For convenience we have multiplied the total action by the coupling constant $8\pi G$, and thus our $\Lambda$ has the mass dimension two and $\Lambda \rightarrow 8\pi G\Lambda$ in the more usual convention.}. The question whether such simple but exotic actions could mimic GR to a sufficient accuracy is outside the scope of the present study.

On the other hand, one could consider whether the $\F(\R)\rightarrow 0$ can be dynamically reached consistently with some general $f(\R)$. Looking at the field equations does not show any apparent problem with this limit. This would realize the ''ground state'' of the metric $\hat{g}_{\mu\nu}$ where all its components vanish, but the connection $\hat{\Gamma}$ can remain well-defined. See Ref.\cite{Banados:2008jx} for interesting discussion of motivations and implications of such a possibility.

\subsection{Coupling matter fields to the connection}
\label{coupl_eq}

Consider matter fields $\Psi$ which couple explicitly to the connection,
$\LA_m=\LA_m(g_{\mu},\hat{g}_{\mu\nu},\Psi)$. The equation of motion for $\R$ retains its form (\ref{r_evol}).
The field equations generalize to
\ba
8\pi G \T_{\mu\nu} & = &
\lb f'(\R) - L \F'(\R)\rb\hat{R}_{\mu\nu} - \frac{1}{2}f(\R)g_{\mu\nu}
+ \F(\R)L_{\mu\nu} \,, \label{geom_b} \\
L_{\mu\nu} & = &
\F(\R)^{\frac{D-4}{2}}\lp {g}_{\mu\nu}{\Box}-{\nabla}_\mu{\nabla}_\nu\rp S(\R,\T)
-\F(\R)^{\frac{D-6}{2}}\lb (D+1)g_{\mu\nu}g^{\al\bt}\F(\R)_{,\al}S(\R,\T)_{,\bt}-2\F(\R)_{(,\mu}S(\R,\T)_{,\nu)}\rb\,.
\label{hateom_b}
\ea
The trace is
\be
L\lp \F(\R)-\F'(\R)\R\rp + f'(\R)\R - \frac{D}{2}f(\R) = 8\pi G \T\,. \label{r_eom_b}
\ee
Here $\T_{\mu\nu}$ is the generalized stress energy tensor
\ba \label{hst}
\T_{\mu\nu}  \equiv
-\frac{2}{\sqrt{-g}}\lb
\frac{\delta\lp\sqrt{-g}\LA_m\rp}{\delta g^{\mu\nu}}
+ \F(\R)\frac{\delta\lp\sqrt{-g}\LA_m\rp}{\delta \hat{g}^{\mu\nu}}
+ \F'(\R)g^{\al\bt}\frac{\delta\lp\sqrt{-g}\LA_m\rp}{\delta \hat{g}^{\al\bt}}\hat{R}_{\mu\nu} \rb\,.
\ea
and $\T$ is its trace $g^{\mu\nu}\T_{\mu\nu}$. In analogy with metric affine gravity \cite{Hehl:1994ue} (MAG), we might call
$\T_{\mu\nu}$  the hyper stress tensor, because it is the sum of the usual stress energy tensor and terms
from variation of the matter lagrangian with respect to the metric that determines the connection. In MAG the
variation of the matter lagrangian with respect to the independent connection is
called the hypermomentum. Despite the same underlying principle of independence of the connection and the metric,
the structure of the present theory is quite different from MAG. We have a considerably simpler system which is
''closer'' to standard GR in the sense that our theory reduces to a metric theory of gravitation with only
$D(D+1)/2+1$ independent field equations. We also note that the system is devoid of the inconsistency related to
projective invariance that plagues MAG (the gravity sector there is invariant under transformations
$\hat{\Gamma}^\al_{\bt\gamma} \rightarrow \delta^\al_\bt V_\gamma$ where $V_\gamma$ is an arbitrary vector, but
the hypermomentum is not). The problem may stem from promoting the connection to a
fundamental degree of freedom, though it is not a tensor field. This interpretation further corroborates our
starting point where the fundamental field is rather the metric associated with the connection than the latter
itself.

To study the conservation laws it is useful to define $\tau_{\mu\nu}$ and its trace
\be
\tau_{\mu\nu} \equiv -2\frac{\delta \LA_m}{\delta \hat{g}^{\mu\nu}}\,, \quad \tau \equiv g^{\mu\nu} \tau_{\mu\nu}\,,
\ee
such that
\be
\T_{\mu\nu}  = T_{\mu\nu} + \F(\R)\tau_{\mu\nu} + \tau\F'(\R)\hat{R}_{\mu\nu}\,.
\ee
The matter action should be invariant under infinitesimal coordinate transformations. With a minimal coupling to
geometry, the stress energy tensor is covariantly conserved. In the case of extended gravity action, this
results in generalized Bianchi identities \cite{Magnano:1993bd,Koivisto:2005yk}. In the present case, i.e.
$\LA_m=\LA_m(g_{\mu},\hat{g}_{\mu\nu},\Psi)$, it is obvious that in general $\nabla_\mu T^{\mu\nu} \neq 0$, and
the conservation laws will have a different form. Consider a coordinate transformation
\be \label{ict}
x^\mu \rightarrow {x'}^{\mu} = x^\mu + \chi^\mu\,.
\ee
The variation of the matter action is
\be \label{var}
\delta_\chi S_m = -8\pi G\int d^D x\sqrt{-g}\lp T_{\mu\nu}\delta_\chi g^{\mu\nu} +
\tau_{\mu\nu}\delta_\chi \hat{g}^{\mu\nu} - 2\frac{\delta \LA_m}{\delta\Psi}\delta_\chi\Psi\rp\,.
\ee
For the two first terms, we have used the definitions of the stress energy tensors. By virtue of the equations
of motion for matter fields, the last term vanishes. It is easy to see by Lie dragging along the vector
$\chi^\mu$ and using a conformal transformation that the two metrics change under (\ref{ict}) as
\ba
\delta_\chi g^{\mu\nu} &= &2\nabla^{(\mu}\chi^{\nu)}\,, \\
\delta_\chi \hat{g}^{\mu\nu} & = & 2\nabla^{(\mu}\chi^{\nu)} + \frac{2}{\F(\R)}\lp
2\chi^{(\mu}\nabla^{\nu)}\F(\R)-g^{\mu\nu}\chi^\al\nabla_\al\F(\R)\rp\,.
\ea
Plugging this into (\ref{var}) gives, after a partial integration
\be
\delta_\chi S_m = 16\pi G\int d^D x\sqrt{-g}\lb
\nabla^\mu \lp T_{\mu\nu} + \tau_{\mu\nu} \rp
- 2\tau_{\mu\nu} \nabla^\mu \log{\F(\R)}
+ \tau \nabla_\nu \log{\F(\R)}
\rb \chi^\nu\,.
\ee
Since this holds for arbitrary $\chi^\mu$ we obtain that
\be
\nabla^\mu \lp T_{\mu\nu} + \tau_{\mu\nu} \rp = 2\tau_{\mu\nu} \nabla^\mu \log{\F(\R)}
+ \tau \nabla_\nu \log{\F(\R)}\,.
\ee
Thus the discrepancy between matter and geometric connections has nontrivial consequences to the equivalence principle.

\section{Stability condition}
\label{stability}

In the following we consider the stability of the theories (\ref{action}) along the lines of Dolgov and Kawasaki \cite{Dolgov:2003px}, who discovered an instability of the scalaron mode in $1/R$-type theories in the weak field limit. This may be called the curvature scalar instability, as it becomes apparent in the equation of motion for the curvature scalar. In our case is given by Eq.(\ref{r_eom}). We are interested in the phenomenological viability of the theories and fix $D=4$. We may expand the functions determining our theories about their GR limits as
\be \label{expand}
f(\R)  =  \R + \epsilon\varphi(\R)\,, \quad \F(\R) = 1 + \epsilon\psi(\R)\,.
\ee
In realistic models compatible with the Solar system experiments, presumably the constant $\epsilon>0$ can be treated as a small parameter.
We yet parameterize the curvature scalar as $\R = -8\pi G T + \epsilon R_1$. Thus $R_1$ measures the deviation from the GR value of $\R$ that is proportional to the trace of the matter (which for simplicity is assumed a perfect fluid here). Then Eq.(\ref{S}) yields
\be \label{S2}
S(\R,T) \approx 1 + \lp\varphi'(\R)-\psi(\R)\rp\epsilon\,.
\ee
Furthermore, we consider weak field regime featuring perturbations $h_{\mu\nu}$ about the flat metric $\eta_{\mu\nu}$ as $g_{\mu\nu} = \eta_{\mu\nu} + \epsilon h_{\mu\nu}$.
Then, up to linear order in $\epsilon$ we obtain by using (\ref{S2}) in Eq.(\ref{lambda}) that
\be
\lambda \approx -3\epsilon\lb \ddot{\varphi}' - \ddot{\psi} - \nabla^2\lp\varphi' - \psi\rp\rb\,,
\ee
where the arguments of $\R$ are suppressed and an overdot means a derivative wrt time. Using Eq.(\ref{r_eom}), one can then obtain the following evolution equation for the perturbation of the curvature scalar:
\ba
\ddot{R}_1 & -& \nabla^2 R_1+ \frac{1}{3\lp\varphi''-\psi'\rp}\lp\frac{1}{\epsilon}-\varphi'\rp R_1 + \frac{2\varphi}{3\lp\varphi''-\psi'\rp} \nonumber \\ & = & 8\pi G \frac{\varphi'''-\psi''}{\varphi''-\psi'}\lb 2\lp \dot{T}\dot{R}_1-(\nabla T)\cdot (\nabla R_1)\rp -8\pi G\lp\dot{T}^2-(\nabla T)^2\rp \rb
 +  8\pi G\lp \frac{T\varphi'}{\varphi''-\psi'}+\ddot{T}-\nabla^2 T\rp\,.
\ea
The right hand side vanishes in the vacuum and is not essential for the curvature scalar instability, which occurs because of the effective mass term. This is the second last term in the left hand side and is typically large since $\epsilon$ is small. Requiring positivity of the effective mass squared and recalling (\ref{expand}) yields the stability condition
\be \label{stab}
f''(\R) \ge \F'(\R)\,.
\ee
The theory is invariant under the change of sign of $\F(\R)$ but the stability condition here isn't simply because we chose to perturb around $\F(\R)=1$. We stress also that this viability criterium applies only around the general relativistic values of these functions.
Let us then look at the implications of this condition in some specific cases.
\begin{itemize}
\item Einstein-Hilbert action $f(\R)=\R-2\Lambda$.
\newline
The viability criterium is simply that the slope of the conformal factor is negative.
\item Metric $f(R)$ gravity.
\newline
Now $\F'(R)=0$ and we have the stability condition $f''(R) \ge 0$. This agrees with the result of \cite{Faraoni:2006sy}, where the analysis of Dolgov and Kawasaki \cite{Dolgov:2003px} in the $1/R$ case was generalized to arbitrary functions $f(R)$.
\item Palatini-$f(\R)$ gravity.
\newline
In this special case $\F(\R)=f'(\R)$, and the criterium (\ref{stab}) is identically satisfied. The absence of the instability in Palatini-$f(\R)$ gravity was discussed in \cite{Sotiriou:2006sf}. As expected, the gravitational perturbation is not propagating but given as a function of the matter content, $R_1=\lp 8\pi G T\varphi' - 2\varphi\rp\epsilon$, to first order in $\epsilon$.
\item Inter- and extrapolating models.
\newline
Consider the parameterization (\ref{alpha}). Now we find that the condition guaranteeing stability is $
f''(\R)(1-\alpha) \ge 0$. Thus the theories that interpolate between metric and Palatini versions of the $f(\R)$ gravities share the same stability criterium with the metric $f(R)$ gravity. Extrapolating beyond the ''Palatini limit'' $\al=1$ where the scalaron is nonpropagating then reverses the stability criterium, and for $\al>1$ one requires $f''(\R)<0$. Note that these conclusions are independent of the detailed form of the interpolating function $\F_\alpha(\R)$ and the criterium is identical when for example $\F_\alpha(\R)=(f'(\R))^\alpha$.
\item Dual theories.
\newline
For the models derivable from dual lagrangians as discussed in section \ref{duality}, the conformal relation is fixed by (\ref{c}) for a given $f(\R)$. The stability condition becomes then
\be
f''(\R)-\frac{1}{\kappa_0\R^2}\lp \R f'(\R)-f(\R)\rp \ge 0\,.
\ee
For the special case $f(\R)=\R-2\Lambda$ in $D=4$ this becomes $\Lambda/(\kappa_0\R) \le 0$. The stability in each region then depends on the sign of the scalar curvature. Given the cosmological constant, we are free to choose $\kappa_0$ of the opposite sign so that positively curved spacetime is stable. (For the cosmological background $\R>0$ unless dominated by a fluid with stiff equation of state $w>1/3$ or violating the null energy condition).

\end{itemize}

\section{On a generalized Eddington's duality}
\label{duality}

The Palatini action for GR in vacuum can be considered as the parent action from which one can derive the Einstein-Hilbert action by eliminating the connection, or the Eddington action by eliminating the metric. Let us consider this in a more general setting and write
\be \label{sum}
S(g,\hat{\Gamma})=\int\lb \sqrt{-g}f(\R) + \sqrt{|\det{\hat{R}_{\mu\nu}}|}h(\R)\rb d^Dx\,.
\ee
We will show that the two terms are equal when $h(\R)=k_0 \R^{\frac{D}{2}}/f(\R)$, where $k_0$ is a dimensionless constant. In the special case $f(\R)=\R-2\Lambda$, the first piece in (\ref{sum}) is well known to be equivalent to the purely metric Einstein-Hilbert action. The vacuum solution $\R=D\Lambda/(D-2)$ can be used in the second piece in (\ref{sum}), which then becomes the Eddington action. Therefore, in this case the two contributions in (\ref{sum}) are easily seen to be the equivalent daughter actions of GR deriving from the Palatini action \cite{Fradkin:1984ai,Banados:2008jx}. In the general case when $f(\R)$ and $h(\R)$ can be nonlinear functions, varying with respect to the metric gives
\be
f'(\R)\hat{R}_{\mu\nu}-\frac{1}{2}f(\R)g_{\mu\nu} + h'(\R)\sqrt{\frac{\det{\hat{R}_{\mu\nu}}}{g}}\hat{R}_{\mu\nu} = 0\,.
\ee
from the trace of this equation we then obtain
\be
\sqrt{\frac{\det{\hat{R}_{\mu\nu}}}{g}}=\frac{\frac{D}{2}f(\R)-f'(\R)\R}{h'(\R)\R}\,.
\ee
 Considering this constraint back in the action (\ref{sum}), it is easy to see that the two pieces transform into each other when $h(\R)=k_0 \R^{\frac{D}{2}}/f(\R)$. 

To make better contact with the present ideas, let us reformulate (\ref{sum}) by beginning from the assumption that there is a metric $\hat{g}_{\mu\nu}$ that generates the independent connection. It is then natural to consider the volume element to be given by this metric instead of its Ricci tensor. We are lead to the bimetric action
\be \label{sum2}
S(g,\hat{g})=\int\lb \sqrt{-g}f(\R) + \sqrt{-\hat{g}}h(\R)\rb d^D x\,.
\ee
In analogy to the above, the trace of the field equation yields the relation between the determinants. This implies that the metrics are now conformally related,
\be
\hat{g}_{\mu\nu} = \lp \frac{\frac{D}{2}f(\R)-f'(\R)\R}{h'(\R)\R} \rp^\frac{2}{D} g_{\mu\nu}         \equiv \F(\R)g_{\mu\nu}\,.
\ee
 It is then obvious that the first part of the action implements the set-up realized in theory (\ref{action}), a function $f(\R)$ of the conformal curvature related to the curvature of the metric by $\F(\R)$.
Again one can deduce that the second piece in (\ref{sum2}) is equal to the first when $h(\R)=(\kappa_0 \R)^{\frac{D}{2}}/f(\R)$, where $\kappa_0$ now has the dimension of mass$^{\frac{2}{D}(4-D)}$. In this case the conformal relation is fixed by the function $f(\R)$ as
\be \label{c}
\F(\R) = \lp \frac{f(\R)}{h(\R)}\rp^\frac{2}{D} = \frac{1}{\kappa_0\R} f^\frac{4}{D}(\R)\,.
\ee
The particular form (\ref{c}) of $\F(\R)$ may thus be of specific interest. In the Einstein-Hilbert case this relation becomes trivial. However, already by including a cosmological constant an interesting deviation from the GR is obtained. In $D=4$ we then have $\F(\R)=(1-2\Lambda/\R)/\kappa_0$. This demonstrates how the duality can naturally generate infrared corrections to gravity due to the inverse relation of $h(\R)$ and $f(\R)$. This kind of inverse curvature-type corrections have been invoked to explain the present acceleration of the universe. (The $f(R)$ type of inverse curvature gravity though fails to produce a viable background in the metric \cite{Amendola:2006kh} and the observed structures in the Palatini formalism \cite{Koivisto:2005yc}. The cosmology of the present theories is left to future studies.) We note that in general the action (\ref{sum2}) is not equivalent to the Palatini-$f(\R)$ even in vacuum. The condition that $\F(\R)$ operates the transformation to the Einstein frame is $\F(\R)=(f'(\R))^\frac{D}{D-2}$. Using the relation (\ref{c}), the differential equation has the solution
\be
f(\R)=\lb \frac{2}{D\kappa_0^{1-\frac{2}{D}}}\lp \R^{2-\frac{D}{2}}- \R_0^{2-\frac{D}{2}} \rp\rb^\frac{-D}{D-4}\,,
\ee
where $\R_0$ is an integration constant. We obtain Palatini-$f(\R)$ gravity only for these specific forms of $f(\R)$. In $D=4$ the solutions are the power-laws
$f(\R) \sim \R^{1/\kappa_0}$. The metric $f(R)$ gravity is recovered when $f(\R) \sim \R^\frac{D}{4}$, which in $D=4$ allows only the Einstein-Hilbert term.

To close this section, we briefly comment relations to bigravity theories. The recently introduced Eddington-Born-Infeld theory \cite{Banados:2008jx,Banados:2010ix} emerges from another type of modification of the Eddington theory, where $\sqrt{|\det{\hat{R}_{\mu\nu}}|}\rightarrow \sqrt{|\det{\hat{R}_{\mu\nu}}|-g_{\mu\nu}}$. This theory can then be shown to belong to the class of bigravity theories \cite{Banados:2008fi} analyzed e.g. in Ref.\cite{Damour:2002ws}. In our case the dynamics of the two metrics is much more constrained, and in fact only one extra degree of freedom, corresponding to the conformal relation between the metrics, is propagating. Some sort of nonlinear bigravity could arise if we considered, in complete symmetry between $g_{\mu\nu}$ and $\hat{g}_{\mu\nu}$, $h$ in (\ref{sum2}) to be a function of $h=h(\hat{g}^ {\mu\nu}R_{\mu\nu})$. However, our starting point prescribes different roles for the metrics, and it is reasonable to consider that this sorts out also the curvature $\R$  in (\ref{r_def}) to mediate the interaction in both parts of the action (\ref{sum2}), though other invariants of course could be constructed from the two metrics.

\section{Conclusions and perspectives}
\label{conclusions}

In Einstein gravity, the spacetime connection is prescribed to be metric compatible, whereas in metric-affine theories one treats the connection as an independent variable. In general the latter results in a completely different theory, which in many instances turns out to be unphysical. In the paper at hand, we considered the possibility that the connection has a prescribed relation to the metric, which however could depend upon the curvature of spacetime. In particular, the connection was assumed to be compatible with the conformal metric $\hat{g}_{\mu\nu} = \F(\R)g_{\mu\nu}$.
It turns out that this subtle adjustment of a foundational principle underlying GR generates a novel type of viable gravitational theories that include both the metric and Palatini gravities as special limits. These C-theories contain but one additional scalar degree of freedom compared to GR, but nevertheless have a remarkably rich structure. We provided several viewpoints into this: the loop formulation (\ref{loop}), the action in the constrained formalism (\ref{action}), the $\F$-frame picture (\ref{hatfr}) and the biscalar-tensor theory (\ref{bist}), which in special cases can be reduced to the Brans-Dicke form (\ref{bd_action}).

The observational implications of the C-theories remain to be studied. The first viability check, stability about Minkowski space, is passed given the condition (\ref{stab}). Severe constraints can certainly be derived from Solar system tests of gravity. It would be very useful to find out how close to unity they force $\F(\R)$ (and how this depends upon $f(\R)$). Cosmological applications then come into question. In particular, one could ask whether C-theories eventually have more to say to the cosmological constant problem or at least to the dark energy problem than the limiting $f(\R)$-theories. In addition, it is natural to consider constraints on the ultraviolet modifications possibly relevant in the early universe. Also the theoretical prospects of C-theories are evidently interesting. It is well known that usual higher-derivative gravity can be renormalizable but only for the unaffordable price of unitarity \cite{Stelle:1976gc}. Whether this is the case for the present class of theories should be studied separately.

Finally, let us note the two obvious generalizations of our starting point (\ref{startingpoint}): one could consider the relation of the metrics to depend upon more general curvature invariants than $\R$, and one could consider the relation to be disformal \cite{Bekenstein:1992pj,Zumalacarregui:2010wj}. These generalizations are in fact intimately related, as it is known that the Palatini variation of theories involving general curvature invariants results in disformal relations \cite{Allemandi:2004wn,Li:2007xw,Olmo:2009xy,Vitagliano:2010pq}. Therefore unifying the Einstein and Palatini versions of generalized higher-derivative actions in the way described here would imply modifying the relation between the metrics from (\ref{startingpoint}). In this exploratory study, our restriction to this form was guided by simplicity and minimality. This choice may also be motivated by special stability properties of actions nonlinear in $\R$ among all possible higher derivative gravities and the special causal structure preserving property of the Weyl rescaling among all possible transformations. However, it would be of interest to study in detail also more general gravity theories within the unified framework.

\section*{Acknowledgments}

We are thankful to Esko Keski-Vakkuri for helpful discussions. L.A. acknowledges support by the DFG through
TRR33 "The Dark Universe". K.E. is supported by the Academy of Finland grants 218322 and 131454.
T.K. is supported by the FOM and the Academy of Finland.

\appendix

\section{Equivalent scalar tensor theories}

\subsection{Einstein frame for $(\xi,\lambda)$}
\label{app_xl}

The lagrangian (\ref{bist}) is transformed into the Einstein conformal frame, denoted by a star, by the conformal transformation $g^*_{\mu\nu}=\xi^\frac{2}{D-2}g_{\mu\nu}$. We obtain
\be
S^* = \int d^D x \sqrt{-g^*}\lb R^* -2\gamma_{ab}(\varphi^c)g_*^{\mu\nu}\varphi^a_{,\mu}\varphi^b_{,\nu} - 4B(\varphi^c) + 16\pi G \xi^\frac{-D}{D-2}\mathcal{L}_m(\Psi,\xi^{\frac{-2}{D-2}}g^*_{\mu\nu})\rb\,.
\ee
We are emplying the notation of Ref.\cite{Damour:1992we} and the fields are denoted as $\varphi^1=\xi$, $\varphi^2=\lambda$. The potential is
\be
B(\xi,\lambda) = \frac{\xi^{\frac{-D}{D-2}}}{4}\lb   \xi\phi(\xi,\ld) - f(\phi(\xi,\ld))\rb\,,
\ee
and the components of the field space metric $\gamma_{ab}(\xi,\lambda)$ defining the nonlinear sigma interaction is
\ba
\gamma_{\xi\xi}(\xi,\lambda) & = & \frac{D-1}{2}\lb \frac{1}{(D-2)\xi^2}-\frac{\F'(\phi)}{\F(\phi)\lp f''(\phi)-\lambda \F''(\phi)\rp \xi} + \frac{(D-2){\F'}^2(\phi)}{4\F^2(\phi)\lp f''(\phi)-\lambda \F''(\phi)\rp^2}\rb\,, \nonumber \\
\gamma_{\xi\lambda}(\xi,\lambda) & = & \gamma_{\lambda\xi}(\xi,\lambda)= \frac{D-1}{4}\lb \frac{(D-2){\F'}^3(\phi)}{2\F^2(\phi)\lp f''(\phi)-\lambda \F''(\phi)\rp^2} - \frac{{\F'}^2(\phi)}{\F(\phi)\lp f''(\phi)-\lambda \F''(\phi)\rp\xi}\rb\,, \nonumber \\
\gamma_{\lambda\lambda}(\xi,\lambda) & = & \frac{(D-1)(D-2){\F'}^4(\phi)}{8\F^2(\phi)\lp f''(\phi)-\lambda \F''(\phi)\rp}\,. \label{gamma}
\ea
Here again $\phi=\phi(\xi,\lambda)$ as given by (\ref{phi1}). However, it turns out that the metric
$\gamma_{ab}(\xi,\lambda)$ is degenerate. One of the eigenvalues of the matrix defined by (\ref{gamma}) vanishes identically,
\ba
\lambda^1_\gamma & = & 0\,, \\ \nonumber
\lambda^2_\gamma & = &
\frac{(D-1)\lb 4 C^2(\phi) (f''(\phi)-\lambda\F''(\phi))^2 -
   4 \F(\phi)(D-2)\F'(\phi) (f''(\phi)-\lambda\F''(\phi)) \xi + (D-2)^2 {\F'}^2(\phi) (1 +
      {\F'}^2(\phi)) \xi^2\rb}{8 \F(\phi)^2 (D-2) (f''(\phi)-\lambda\F''(\phi))^2 \xi^2}\,.
\ea
Therefore we cannot invert $\gamma_{ab}(\xi,\lambda)$ and straightforwardly implement the results of \cite{Damour:1992we} to analyze the PPN limit. Does this imply that there is only one additional propagating scalar degree of freedom compared to GR? That would agree with the following section where we find that the scalar $\lambda$ can be eliminated in terms of $\R$ and $T$, while $\R$ needs, in general, to be solved from a dynamical equation. The eigenmodes corresponding to $\lambda^1_\gamma$ and $\lambda^2_\gamma$ are, respectively
\ba
d\tilde{\varphi}_1 & = & \frac{-{\F'}^2(\phi)\xi}{\lambda\F(\phi)\F''(\phi)-\F(\phi)f''(\phi)+\F'(\phi)\xi}d\xi + d\lambda\,, \\
d\tilde{\varphi}_2 & = & \frac{\lambda\F(\phi)\F''(\phi)-\F(\phi)f''(\phi)+\F'(\phi)\xi}{{\F'}^2(\phi)\xi}d\xi + d\lambda\,.
\ea
in $D=4$.

\subsection{Special case: exponential $\F$}
\label{exponential}

To proceed, we use the pair $(\xi,\phi)$ instead of $(\xi,\lambda)$ in (\ref{bist}) and onwards.
Let us consider the case of exponential relation $\F(\R)=\F_0 e^{2k\R/(D-2)}$ where $\F_0$ are $k$ some constants,
the scaling $\F_0$ being irrelevant and $k$ having the dimension one per mass squared.
Now (\ref{varphi1},\ref{varphi2}) can be solved by
\be
\tilde{\varphi}^{(1)}  = \phi+\frac{k\xi^2}{2}\,, \quad  \tilde{\varphi}^{(2)}
= \phi - \frac{\log{\frac{\xi}{\xi_0}}}{k}\,.
\ee
The inverse transformation is, assuming $\xi>0$,
\ba
\phi  =  \vpu - \frac{W\lp\lb k\xi_0e^{k(\vpu-\vpd)}\rb^2\rp}{2k}\,, \quad
\xi = \frac{W^\frac{1}{2}\lp\lb k\xi_0e^{k(\vpu-\vpd)}\rb^2\rp}{|k|}\,,
\ea
where $W(x)$ is the Lambert $W$-function which solves the equation $x=We^W$. In the following we denote it
just $W(\vp^c)$ since the argument of the function will have the fixed form $x=k^2\xi^2_0e^{2k(\vpu-\vpd)}$ as
above. In terms of the new fields, the action becomes
\be \label{st}
S^* = \int d^D x \sqrt{-g^*}\lb R^* - \frac{D-1}{(D-2)k^2}\lp\partial \vpd\rp^2
- 4B(\vp^c)\rb + 16\pi G S_m(\Psi,A^2(\vp^c)g^*_{\mu\nu})\,,
\ee
where the two functions are
\ba
B(\vp^c)                & = & \frac{1}{4}\lp\frac{|k|\xi_0}{W^\frac{1}{2}(\vp^c)}\rp^\frac{D}{D-2}
\lb \frac{W^\frac{1}{2}(\vp^c)}{|k|\xi_0}\lp\vpu-\frac{W(\vp^c)}{2k}\rp
+f\lp \vpu-\frac{W(\vp^c)}{2k}\rp\rb\,, \\
A(\vp^c) & = & \lp\frac{|k|}{W^\frac{1}{2}(\vp^c)}\rp^\frac{1}{D-2}\,.
\ea
Varying the action (\ref{st}) with respect to the auxiliary field $\vpu$ now yields, when $W(\vp^c) \neq -1$,
\ba
  \frac{D-2}{2}\lp \frac{|k|\xi_0}{W^\frac{1}{2}(\vp^c)}\rp^\frac{D}{D-2}\lb
\lp \vpu-\frac{W(\vp^c)}{2k}+\frac{1}{|k|}\rp \frac{W^\frac{1}{2}(\vp^c)}{\xi_0}
+ f'\lp \vpu-\frac{W(\vp^c)}{2k}\rp\rb   -  2D|k|\xi_0 B(\vp^c) \nonumber \\
 =  - 8\pi G|k|T\,.
\ea
Let us consider a special case. We choose the Einstein-Hilbert form $f(\R)=\R$ in $D=4$ dimensional vacuum $T=0$. The theory then assumes the form
\be
S = \int d^4x\sqrt{-g} \lb \phi R -\frac{3 \lp 8 - \phi (7 - 2\phi)\rp}{2(\phi-2)^4}\lp\partial\phi\rp^2 - \frac{(\phi-1)^2}{k(\phi-2)}\rb\,.
\ee
 This illustrates the facts that even linear $f(\R)$ result in new gravitational effects in the C-theory context and that we cannot reduce the models in general to the pure Brans-Dicke form, but the coupling is instead nonlinear. Here, in the limit $k\rightarrow 0$, the field $\phi$ is fixed to $\phi=1$ and Einstein gravity is recovered.

\subsection{A subtlety}
\label{nonlocal}

As noted in section (\ref{bst}), we can rewrite the action (\ref{action}) as
\be
\mathcal{L}_g = f(\phi) + \hat{\ld} - \F(\phi)\ld + \xi(\R-\phi)\,.
\ee
Now solving the constraint equation for $\lambda^{\mu\nu}$ which is $\hat{g}_{\mu\nu}=\F(\phi)g_{\mu\nu}$ and then plugging the constraint imposed by $\xi$ back into the lagrangian give
\ba \label{wrong}
\mathcal{L}_g & = & f(\phi)+\xi\lp R-\frac{(D-1)(D-6)}{4\F^2(\phi)}\lp\partial \F(\phi)\rp^2-\frac{D-1}{\F(\phi)}\Box \F(\phi) - \phi \rp \nonumber \\
& = & f\lp R-\frac{(D-1)(D-6)}{4\F^2(\phi)}\lp\partial \F(\phi)\rp^2-\frac{D-1}{\F(\phi)}\Box \F(\phi)\rp\,.
\ea
One implication of this would be that when $f(\R)=\R$ the theory reduces to GR + a minimally coupled scalar field. However, we emphasize that the theory on the second line of (\ref{wrong}) is not in general the same as in the first one. This is because the second equality in (\ref{wrong}) is obtained by plugging a dynamical equation of motion back into the action, which in general is not legitimate. This is obvious when considering a simple scalar theory $2\mathcal{L}_\varphi = \varphi\lp\Box-m^2\rp\varphi$: if we plug the Klein-Gordon equation $\Box\varphi=m^2\varphi$ back into the action it vanishes. The second line in (\ref{wrong}) should be regarded as the action for a fixed configuration of $\phi$ that is a solution to the constraint equation, with the $\phi$ appearing there not a variational degree of freedom.

\bibliography{conformal}

\end{document}